\documentclass[prb,twocolumn,aps,showpacs]{revtex4}
\usepackage{epsfig}
\usepackage{bm}
\begin{document}
\date{\today}
\title{Ground-state fidelity in the BCS-BEC crossover}
\author{Ayan Khan and Pierbiagio Pieri}
\affiliation{\mbox{Dipartimento di Fisica, Universit\`{a} di Camerino, I-62032 Camerino, Italy}}
\begin{abstract}
The ground-state fidelity has been introduced recently as a tool to investigate quantum phase transitions. Here, we apply this concept in the context of a crossover 
problem. Specifically, we calculate the fidelity susceptibility for the BCS ground-state wave function, when the intensity of the fermionic attraction is varied from weak
to strong in an interacting Fermi system, through the BCS-BEC crossover. 
Results are presented for contact and finite-range attractive potentials and for both continuum and lattice models.
We conclude that the fidelity susceptibility can be useful also in the context of crossover problems. 
\end{abstract}

\maketitle
\section{Introduction}
A quantum phase transition is an abrupt change of the ground state of a many-body system when a controlling parameter $\lambda$ of the Hamiltonian crosses a critical
value $\lambda_c$. It is then quite natural to expect~\cite{zanna} that the overlap $F(\lambda+\delta\lambda,\lambda)\equiv |\langle\Psi(\lambda+\delta\lambda) | \Psi(\lambda)\rangle|$ between the ground states corresponding to two 
slightly different values of the parameter $\lambda$, should manifest an abrupt drop when the small variation $\delta\lambda$ crosses 
$\lambda_c$. Such overlap, which has been named ``ground-state fidelity'' in the literature, should thus provide a tracer of a quantum phase transition.
An appealing feature of the ground-state fidelity is that it does not rely on the explicit knowledge, or even the very existence, of an order parameter associated with the 
quantum phase transition.

The ground-state fidelity approach to quantum phase-transitions has been
tested in a variety of models\cite{zanna,You07,Yang07,chen07,campos07,zan07,cozzini07,buonsante07,hamma08,tzeng08,chen08,gu08,abasto08,paunkovic08,Zhou08,zha09,wan09,eri09}, and has been found 
to be quite effective in signalling the presence of a quantum phase transition. 
In particular, even phase transitions which are not associated directly with a local order parameter, like transitions topological in nature or of the 
Beresinskii-Kosterlitz-Thouless type have been found to be detectable by the ground-state fidelity (or derived quantities like the fidelity susceptibility)\cite{Yang07,hamma08,abasto08,garnerone09}. [Some phase transitions of high order in terms of the controlling parameter $\lambda$ have been found, however, to escape the analysis based 
on the ground-state fidelity\cite{campos07}.]

In the present paper we make a step further in the study of the potentialities of the ground-state fidelity, by analyzing its applicability beyond the 
domain for which it was originally conceived. We consider specifically a 
{\em crossover\/} problem, that is, a many-body system for which 
a substantial change in the nature of its ground state occurs over a {\em finite} range of the controlling parameter $\lambda$,
rather than abruptly at a critical value $\lambda_c$.    
On physical grounds, we expect that the sudden drop of the ground-state fidelity at a critical value $\lambda_c$ should in this case be replaced by a minimum, located in the region of $\lambda$ where the ground state is changing more rapidly, namely, in the crossover region. In terms of the fidelity susceptibility (defined shortly 
below), the divergence at a critical point should correspondingly be replaced by a peak, whose width should be 
associated with the width of the crossover region.

To test the above ideas we consider specifically the BCS-BEC crossover, 
namely, the evolution of the ground state of a fermionic system in the presence
of an attractive potential which is progressively increased in its strength. 
This evolution, first studied in Refs.~\onlinecite{eagles,legget,NSR}, has 
been studied then quite extensively, firstly in connection with the physics of 
high-temperature cuprate superconductors\cite{randeria,micnas,haussmann,pistolesi,zwerger,levin}, and more recently 
with ultracold Fermi atoms, where this crossover has been realized experimentally\cite{ketterle,regal,chin,grimm,stan,jin}. Several models 
supporting this crossover will be analyzed, both in continuum and discrete space. Such a systematic analysis will allow us to draw conclusions of sufficient generality,
which we believe to be applicable also to other crossover problems.

The paper is organized as follows. In Sec.~II we introduce the fidelity 
susceptibility and derive its expression for the BCS wave-function. This 
quantity is calculated explicitly in Sec.~III for the 
BCS-BEC crossover in continuuum models. We consider 
specifically the three-dimensional contact and Nozier\'es-Schmitt-Rink 
potentials, and compare the information extracted from the fidelity 
susceptibilty with what is already known in the literature for the
BCS-BEC crossover in these models.
A similar analysis is presented in Sec.~IV for the attractive Hubbard 
model (in both two and three dimensions).
Section V gives our conclusions.

\section{The fidelity susceptibility for the BCS wave function}
The ground-state fidelity $F(\lambda+\delta\lambda,\lambda)$ depends, by its definition, on both the controlling 
parameter $\lambda$ and its variation, $\delta\lambda$. The somewhat 
artificial dependence on the actual value of the parameter 
$\delta\lambda$ can be eliminated by considering 
the limiting expression for the ground-state fidelity when  $\delta\lambda$ approaches zero. For small $\delta\lambda$:  
\begin{eqnarray}
F(\lambda+\delta\lambda,\lambda)^2&=&\left[\langle\Psi (\lambda)|+
\delta\lambda \, \frac{\partial \langle\Psi (\lambda)|}{\partial\lambda}  \nonumber\right.\\
 &+&  \left. \frac{(\delta\lambda)^2}{2} \frac{\partial^2 \langle\Psi(\lambda)|}{\partial{\lambda}^2} 
\right]\cdot |\Psi(\lambda)\rangle\\
 &=&1-\frac{(\delta\lambda)^2}{2}\frac{\partial\langle\Psi 
(\lambda)|}{\partial\lambda}\cdot \frac{\partial
|\Psi(\lambda)\rangle}{\partial \lambda},
\end{eqnarray} 
 where the state $|\Psi(\lambda)\rangle$ is assumed to be real and normalized.
A sudden drop of the ground-state fidelity at the critical point will then 
correspond to a divergence of the {\em fidelity susceptibility}\cite{You07,Yang07}:
\begin{eqnarray}
\chi(\lambda)&\equiv& - \frac{1}{\Omega}\lim_{\delta\lambda\to 0} \frac{4 \ln F(\lambda+\delta\lambda,\lambda)}{(\delta \lambda)^2}\label{chi1}\\
&=&\frac{1}{\Omega}\frac{\partial\langle\Psi 
(\lambda)|}{\partial\lambda}\cdot \frac{\partial
|\Psi(\lambda)\rangle}{\partial \lambda} \label{chi2}. 
\end{eqnarray}
Note that, in order to deal with a meaningful quantity in the thermodynamic limit, the expressions on the right-hand side of Eqs.~(\ref{chi1}) and (\ref{chi2}) in the 
definition of the fidelity susceptibility have been divided by the system volume $\Omega$. 
For a sufficiently large system, one has in fact  $\ln F(\lambda,\lambda')\propto \Omega$. Indeed, barring the case when a correlation lenght 
$\xi$ diverges (i.e. when $\lambda$ or $\lambda'$  sit exactly at a critical point), the system can be thought as a collection (tensor product) of 
many identical subsystems of volume $L^D$ (with $L\gg\xi$ and where $D$ is the spatial dimension).
 The overlap between two different ground states 
$|\Psi(\lambda)\rangle$ and $|\Psi(\lambda')\rangle$ will 
be then the product of the overlaps in each individual 
subsytem, yielding $F(\lambda,\lambda')\propto 
f(\lambda,\lambda')^{N_s}$, where $f(\lambda,\lambda')$ is 
the overlap in each subsystem and $N_s=\Omega/(L^D)$ is the total number of subsystems. 
This implies then $\ln F(\lambda,\lambda')\propto \Omega$ except when $\lambda$ or $\lambda'$ are exactly at a critical point\cite{campos07}.
 
In this paper, we are interested in calculating the fidelity susceptibility 
$\chi(\lambda)$ across the BCS-BEC crossover.
Previous studies~\cite{legget,NSR,per04} have shown that the BCS wave-function
provides a reasonably good approximation for the ground-state wave-function
over the whole BCS-BEC crossover, from 
the weak-coupling limit of highly overlapping Cooper pairs to the 
strong-coupling limit of tightly bound dilute composite bosons.

 We will thus
calculate $\chi(\lambda)$ for the BCS wave-function:
\begin{equation}
|\Psi(\lambda)\rangle=\prod_{{\bf k}}[u_{k}(\lambda)+v_{k}(\lambda)c_{{\bf k}\uparrow}^{\dagger}c_{-{\bf k}\downarrow}^{\dagger}]|0\rangle\; .
\label{BCS}
\end{equation}
Here, $c_{{\bf k}\sigma}^{\dagger}$ creates a fermion in the single-particle state of wave-vector ${\bf k}$, spin $\sigma$ and energy $\epsilon_k$, $u_k$ and $v_k$ are the usual BCS coherence factor $v_k^2=1-u_k^2=(1-\xi_k/E_k)/2$, with 
$\xi_k=\epsilon_k-\mu$, $E_k=\sqrt{\xi_k^2+\Delta_k^2}$, where $\mu$ is the 
chemical potential and $\Delta_k$ the BCS gap function.  
The parameter $\lambda$ in Eq.(\ref{BCS}) stands generically for the appropriate coupling strength of the attractive interaction  $V_{\lambda}(k,k')$ which is driving the BCS-BEC 
crossover. The dependence of $u_k$ and $v_k$ on $\lambda$ in Eq.(\ref{BCS}) is 
determined by the gap function $\Delta_k$ and chemical potential $\mu$, which depend on $\lambda$ through the gap and particle number equations:
\begin{eqnarray}
\Delta_{k}&=&-\int\!\!\frac{d {\bf k}'}{(2\pi)^D} \, V_{\lambda}({\bf k},{\bf k}')\frac{\Delta_{k'}}{2E_{k'}}\label{defdelta}\\
n&=&\int\!\!\frac{d{\bf k}}{(2\pi)^D} \, 2\, v_k^2 \label{defden}
\end{eqnarray} 
where $n$ is the particle number density.

When the BCS wave function is inserted in Eq.~(\ref{chi2}) for $\chi(\lambda)$ 
one obtains:
\begin{equation}
\chi(\lambda)=\int\!\!\frac{d {\bf k}}{(2\pi)^{D}} \left[\left(\frac{d u_{k}}{d\lambda}\right)^2+ \left(\frac{d v_{k}}{d\lambda}\right)^2\right]
\end{equation}
which, after some manipulations, can be written
\begin{equation}
\chi(\lambda)=\int\!\!\frac{d {\bf k}}{(2\pi)^{D}} \frac{1}{4E_{k}^4}\left[\Delta_{k}\frac{d\mu}{d\lambda}+\xi_{k}\frac{d\Delta_{k}}{d\lambda}\right]^2
\, .\label{chiBCS}
\end{equation}
In the next two sections we will analyze the behaviour of the fidelity susceptibility for several type of attractive interaction $V_{\lambda}(k,k')$, by solving the coupled Eqs.~(\ref{defdelta}) 
and~(\ref{defden}) and by calculating then 
$\chi(\lambda)$ as determined by Eq.~(\ref{chiBCS}). Section III will consider 
continuum models, while section IV will deal with a lattice model (the attractive Hubbard model). In that case the integration over $k$ in  Eqs.~(\ref{defdelta}),(\ref{defden}) and ~(\ref{chiBCS}) will be limited to the first Brillouin zone.

\section{Continuum models: contact and finite-range potentials}
The simplest model Hamiltonian for the BCS-BEC crossover describes a system of fermions of mass $m$ in continuum space, mutually interacting via a contact 
($\delta$-like) interaction:  
\begin{eqnarray}\label{contact}
H&=&\sum_{\sigma}\int\!\! d\mathbf{r} \, \psi_{\sigma}^{\dagger}(\mathbf{r}) \frac{-\nabla^2}{2m}\psi_{\sigma}(\mathbf{r})\nonumber\\
&&+ \, g\int \!\! d\mathbf{r} \, \psi_{\uparrow}^{\dagger}(\mathbf{r})\psi_{\downarrow}^{\dagger}(\mathbf{r}) \psi_{\downarrow}(\mathbf{r})\psi_{\uparrow}(\mathbf{r})\, .
\end{eqnarray}
The Hamiltonian (\ref{contact}) leads to ultraviolet divergencies, as it can be seen in the gap equation (\ref{defdelta}) when both the gap function and the interaction 
do not depend on wave vector (as it occurs for a contact potential). These divergencies are, however, eliminated by expressing physical quantities in terms of 
the two-body scattering length $a_F$ rather than the bare coupling $g$.

The above Hamiltonian (with the appropriate ultraviolet regularization) has been studied quite extensively in the context of the
BCS-BEC crossover, especially after the advent of experiments with ultracold Fermi atoms in the presence of a Fano-Feshbach resonance\cite{ketterle,regal,chin,grimm,stan,jin}.
For these systems the Hamiltonian (\ref{contact}) can in fact be derived from first principles\cite{simonucci}, as the effective Hamiltonian describing the 
physics of the relevant degrees of freedom close to the Fano-Feshbach resonance.

For a  three dimensional contact potential the 
gap equation (\ref{defdelta}), when expressed in terms of the scattering length 
$a_F$ reads: 
\begin{equation}
-\frac{m}{4\pi a_F}=\int\!\!\frac{d {\bf k}}{(2\pi)^3} \left( \frac{1}{2 E_k}
-\frac{m}{k^2} \right)\;.
\label{gapcont3D}
\end{equation}
The coupled Eqs.~(\ref{gapcont3D}) and (\ref{defden}) determine the gap $\Delta$ and chemical potential $\mu$ in terms of the scattering length $a_F$ or, better, of the dimensionless coupling parameter $(k_{F} a_{F})^{-1}$, which is normally used as the coupling strength parameter for the 3D contact potential 
(here $k_F\equiv(3\pi^2 n)^{1/3}$, such that the scattering length $a_F$ is compared with the average interparticle spacing $k_F^{-1}$).

In terms of this parameter the BCS and BEC limits correspond in principle to the conditions $(k_{F} a_{F})^{-1}\ll -1$ and $(k_{F} a_{F})^{-1} \gg 1$, respectively. 
Previous studies\cite{ran93,hau94,marini,per02,pie04} have 
shown, however, that the crossover between the above two different physical situations is limited in practice to the rather narrow region $-1 \lesssim (k_{F} a_{F})^{-1}\lesssim 1$.

\begin{figure}[t]
\begin{center}
\epsfxsize=8cm
\epsfbox{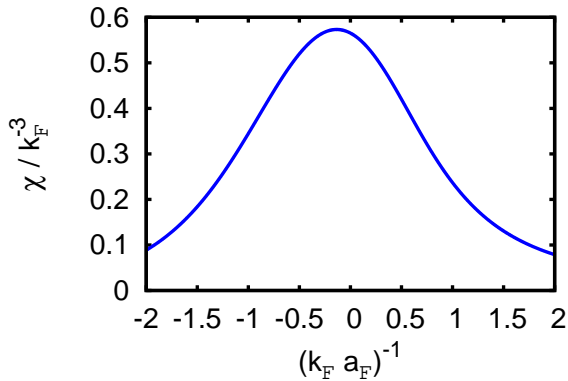}
\vspace{0.5cm}
\caption{Fidelity susceptibility (in units of $k_F^{-3}$) for the three-dimensional contact potential as
a function of the dimensionless coupling strength $(k_F a_F)^{-1}$.}
\end{center}
\label{3dc}
\end{figure}

We have calculated the fidelity susceptibility (\ref{chiBCS}), where as a controlling parameter $\lambda$  we 
have taken the dimensionless coupling strength (i.e. $\lambda=(k_{F} a_{F})^{-1}$), by solving the coupled Eqs.~(\ref{gapcont3D}) and (\ref{defden}) 
(which can be suitably expressed in terms of
elliptic integrals\cite{marini}) to determine $\Delta(\lambda)$,
$\mu(\lambda)$ and then $\chi(\lambda)$ through the BCS-BEC crossover.
The resulting fidelity susceptibility  $\chi((k_{F} a_{F})^{-1})$ is presented in Fig.~1.

The fidelity susceptibility presents a rather symmetric peak, which is located in the middle of the crossover region (specifically, the peak 
position corresponds to $(k_{F} a_{F})^{-1}\simeq -0.14$) and whose half-height width marks precisely the borders of the crossover region $-1 \lesssim (k_{F} a_{F})^{-1}\lesssim 1$ mentioned above. Note that the size of the crossover region for 
the three-dimensional contact potential was determined in previous studies by 
calculating specific physical quantitities, like e.g. the chemical potential, 
the BCS gap, or the superfluid critical temperature and by comparing their 
numerical values with analytic expressions valid in the BCS and BEC limits, 
respectively. 
This empirical way of defining the range of the crossover region, even though physically sound, could be criticized because of some degree of arbitrariness in choosing the physical quantity to look after, or the quantitative 
criterion to conclude that a specific asymptotic (BEC or BCS) expression has been 
effectively reached.

The plot of the fidelity susceptibility $\chi((k_{F} a_{F})^{-1})$ of Fig.~1 provides 
a somewhat more ``intrinsic'' criterion to locate the position and width of
the crossover region, since it is not based on a specific physical quantity but
on the measure of the rapidity of change of the ground-state 
wave-function through the crossover. The agreement between the position and width of the crossover region, 
as determined by the fidelity susceptibility, with previous results obtained with more empirical criteria to characterize the 
crossover region corroborates these previous results while proving, at the 
same time, the utility of the ground-state fidelity for studying also 
crossover problems.  

We pass now to consider a finite-range potential. Specifically, we consider the separable potential introduced by Nozi\`eres and Schmitt-Rink\cite{NSR} (NSR):
\begin{equation}
V(k,k')=  \frac{-V}{\sqrt{1+(k/k_0)^2}\sqrt{1+(k'/k_0)^2}}
\end{equation}
where $k_0$ sets the range of the potential in momentum space. The finite range of the NSR potential allows for the occurrence of the 
density induced BCS-BEC crossover~\cite{and99} which is instead not possible in the case of a contact 
potential.

We have calculated the fidelity susceptibility for the NSR potential (with 
$\lambda=V$) for various values of the particle density (which can be parametrized 
by the ratio between $k_F$ and the momentum range of the potential $k_0$). For each 
density, our calculated $\chi(V)$ are peaked at a value $V=V_p$ whose position indicates where 
the rate of change of the BCS wave-function with respect to $V$ is maximal. 
As already argued above, the value $V_p$ should thus be located in the crossover region.    

\begin{figure}[t]
\begin{center}
\epsfxsize=8cm
\epsfbox{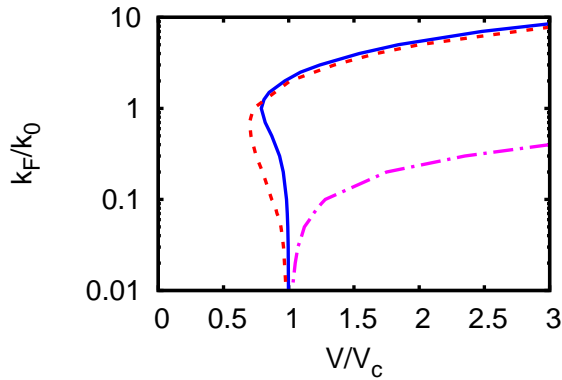}
\vspace{0.5cm}
\caption{Crossover ``phase diagram'' for the three-dimensional NSR potential. 
The full curve was obtained by determining the position of the peak in the fidelity susceptibility $\chi(V)$ at given values of the parameter $k_F/k_0$. 
The dashed and dash-dotted curves correspond to the BCS and BEC borders of the crossover region, respectively, as determined in Ref.~\onlinecite{and99}. 
The potential strength $V$ is in units of the critical potential strength $V_{{\rm c}}= 4 \pi/(m k_0)$.}
\end{center}
\end{figure}

Fig.~2 compares the peak position $V_p$ extracted from the fidelity susceptibility for several values of $k_F/k_0$ (full line) with the curves 
defined by the conditions $k_F \xi_{{\rm pair}}= 2 \pi$ (dashed line)  and $k_F \xi_{{\rm pair}}= 1/\pi$ (dash-dotted line), obtained previously  
in Ref.~\onlinecite{and99} and introduced there to characterize the width of the crossover region (thus defining a sort of ``phase diagram'' for the
BCS-BEC crossover). 
In particular, $\xi_{{\rm pair}}$ represents the pair correlation length, as defined in Ref.~\onlinecite{pistolesi}, and provides 
an estimate of the Cooper pair radius.
The BCS region is characterized by large overlapping Cooper pairs (such that $k_F \xi_{{\rm pair}}\gg 1$) while for the BEC region, 
with small nonoverlapping boson, $k_F\xi_{{\rm pair}}\ll 1$.
The two values ($2 \pi,1/ \pi$) of the parameter $k_{F} \xi_{{\rm pair}}$ 
were taken in Ref.~\onlinecite{pistolesi} as representative of the BCS and BEC borders of the crossover region,
essentially on the basis of the behaviour of the chemical potential as a function of the parameter $k_F \xi_{{\rm pair}}$ itself.

We can see from Fig.~2 that the curve corresponding to the peak position $V_p$ extracted from the fidelity susceptibility lies as expected in the middle 
of the crossover region for 
$k_F/k_0\lesssim 1$. For higher densities it approaches instead the BCS edge of this region.  This may be due to the fact that at these densities and for  
$k_F\xi_{{\rm pair}}$ of order unity (as in the middle of the crossover region), the range of the attractive potential becomes larger than the Cooper pair size. 
This favours the clustering of pairs (and eventually leads to an instability for sufficiently strong attraction, because of dominant pair-pair 
attractive interaction\cite{micnas,pis96,Ropke}).
The pair correlation length will thus be increasingly influenced by inter-pair correlations rather than by intra-pair correlations. This implies that at high
densities $\xi_{{\rm pair}}$ tend to overestimate the actual radius of a Cooper pair. 
At high densities the curve obtained from the peak position in the fidelity susceptibility, which is not based on $\xi_{{\rm pair}}$ and is thus not influenced by this
effect, is thus arguably a better indicator for the position of the crossover region than  what is obtained from the calculation of $k_F\xi_{{\rm pair}}$ itself.

Note finally the merging of the three curves into the single point $V=V_c$ for 
vanishing $k_F/k_0$. This is expected on physical grounds since in the two-body problem
a qualitative change in the ground state (from a delocalized to a localized wave function) occurs precisely 
at $V=V_c$. It is indeed easy to verify that the fidelity susceptibility calculated over the two body wave function 
diverges at $V=V_c$. At low densities, the fidelity susceptibility approaches the two-body behaviour and is thus peaked aroud 
$V_c$.
More generally, when the interparticle distance gets larger than the range of the potential the universal behaviour described
by the contact potential is recovered. Since, as we have seen above, for this
potential the crossover occurs in the region $|k_F a_F|\gtrsim 1$, it is then 
clear that when $k_F$ is vanishing the scattering length $a_F$ is bound to diverge in order to 
keep the product  $|k_F a_F|\gtrsim 1$, thus pinning the crossover region close to $V=V_c$ where the scattering length diverges.

\section{Attractive Hubbard model}

We analyze finally the fidelity susceptibility calculated over the BCS wave function for the attractive Hubbard model with nearest-neigbour hopping:
\begin{equation}
H=-t \sum_{\langle i,j \rangle \sigma} c^{\dagger}_{i\sigma}c_{j\sigma} - U \sum_i c^{\dagger}_{i\uparrow}c_{i\uparrow}c^{\dagger}_{i\downarrow}c_{i\downarrow}
\end{equation}
where $\langle i,j \rangle$ indicates a sum over nearest-neighbor pairs and $c^{\dagger}_{i\sigma}$ creates one electron with spin $\sigma$ in the Wannier state centered around the lattice site $i$.
For an $s$-wave gap, which does not depend on ${\bf k}$, the wave vector sums in the gap and particle number Eqs.~(\ref{defdelta}),(\ref{defden}) and in Eq.~(\ref{chiBCS}) for the fidelity
susceptibility are more efficiently calculated by converting them into integrals over the energy and by using the density of states appropriate for the lattice kinetic energy dispersion. 

\begin{figure}[t]
\begin{center}
\epsfxsize=8cm
\epsfbox{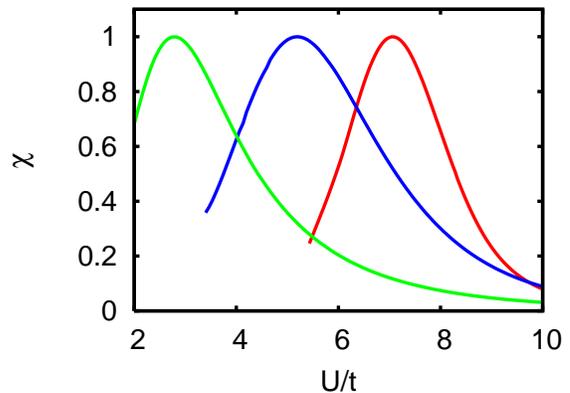}
\vspace{0.5cm}
\caption{Fidelity susceptibility (normalized for convenience to the peak 
value) for the three-dimensional attractive Hubbard model at filling values 
$n=0.25, 0.05, 0.005$ from left to right.}
\end{center}
\end{figure}

\begin{figure}[ht]
\begin{center}
\epsfxsize=8cm
\epsfbox{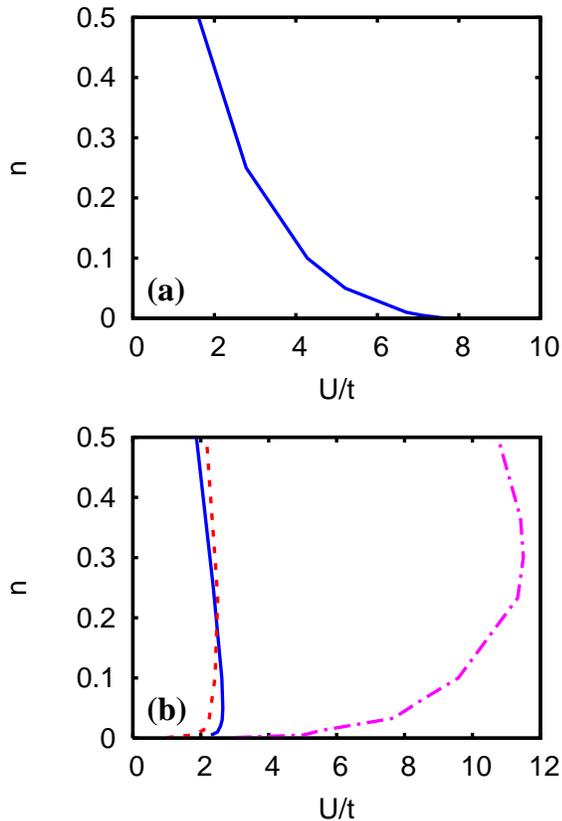}
\vspace{0.5cm}
\caption{Attractive Hubbard model: position of the peak in the fidelity 
susceptibility $\chi(U)$ in the plane $(U/t,n)$ in three (a) and two (b) 
dimension. In two dimensions the curves corresponding to the BCS and BEC 
borders of the crossover region as determined in  Ref.~\onlinecite{and99}
are also reported for comparison (dashed and dash-dotted curves, respectively).}
\end{center}
\end{figure}

The resulting fidelity susceptibility for the three-dimensional attractive 
Hubbard model is shown in Fig.~3, at three 
different values of the filling factor $n$ (particle number per lattice site). For decreasing filling, the peak in the fidelity 
susceptibility progressively narrorws in width, while its position approaches the critical value ($U_c\simeq 7.9 t$) above which  
a bound state appears in the two-body problem.
The width of the crossover region as extracted from the half-height width in the fidelity susceptibility thus  
shrinks for decreasing density. This overall behaviour of the fidelity susceptibility suggests a progressive approaching of the BCS-BEC crossover 
towards a quantum phase transition for decreasing density.
It is in fact the presence of a finite density in the many-body problem which smears out over a finite coupling range the sudden change occurring 
at $U_c$ in the two-body problem.
In addition, dynamical mean field calculations\cite{metzner,capone} have shown the occurrence of a quantum phase transition  between a Fermi liquid state 
and a paired state when superconductivity is 
artifically suppressed in the attractive Hubbard model. This transition, 
which also approaches the critical value $U_c$ for the two-body problem for vanishing density, is  transformed into a crossover 
when superconducting correlations are restored (in the very same way the metal-insulator transition is transformed into a crossover 
within the antiferromagnetic state in
the half-filled repulsive Hubbard model). The behaviour of the fidelity 
susceptibility at low densities can thus be interpreted as the remnant, 
within the superconducting state, of such underlying  quantum phase transition
in the normal state.  
             
Fig.~4 compares finally the peak position in the fidelity susceptibility $\chi(U)$  in three and two  
dimensions (full lines in panel (a) and (b), respectively). In two dimensions the
curves corresponding to the conditions $k_F \xi_{{\rm pair}}= 2 \pi$ (dashed line)  and $k_F \xi_{{\rm pair}}= 1/\pi$ (dash-dotted line), obtained previously  
in Ref.~\onlinecite{and99} are also reported (the three dimensional Hubbard model was instead not considered in Ref.~\onlinecite{and99}).  
From Fig.~4~(b) we can see that at low filling the peak position in $\chi(U)$ 
lies in the middle of the crossover region, while at higher filling it bends 
towards the BCS border of the crossover region, similarly to what found for the three-dimensional NSR continuum potential. We think that also in this case 
the increasing importance of inter-pair correlations at higher densities 
explains the relative behaviour of the two curves in this regime. Note, in this respect, 
that for the Hubbard model it is the lattice spacing to provide the additional
length scale relevant at high densities, playing the same role as the finite 
range of the interaction for the NSR potential.

Note, finally, that in two dimensions the three curves reported in Fig.~4~(b) tend to 
the ``critical'' value for  the two-body problem ($U=0$) only at extremely 
low fillings (recall that in two dimensions, a bound state occurs 
in the two-body problem as soon as the attraction $U$ is switched on), while  
in  three-dimensions (Fig.~4~(a)) the peak position in the fidelity approaches the two-body critical 
value $U_c\simeq 7.9 t$ more progressively.
The behaviour of the three curves at low fillings can be explained by recalling that in two dimensions the binding energy of the two body problem
$\epsilon_0 \propto \exp(-t/U)$, while at low densities the Fermi energy relative to the bottom of the band  is proportional to the filling ($\epsilon_F \propto n$), 
such that when $\epsilon_0$ and $\epsilon_F$ are of the same order (as it occurs for a crossover curve) one has $n\propto\exp(-t/U)$. This explains the rapid collapsing to zero of
the three curves when $U/t \lesssim 1$. 
   
\section{Concluding remarks}  
In this paper, we have tested the usefulness of the fidelity susceptibility in the context of a crossover problem.  
We have considered specifically the BCS-BEC crossover, for which the BCS wave function is known to provide 
a reliable description of the ground-state properties, and calculated the fidelity susceptibility over the BCS wave function 
for several models exhibiting the above crossover. For the  
three-dimensional contact potential, we have found that the peak position in the fidelity susceptibility and its width in terms of the dimensionless 
coupling parameter $(k_F a_F)^{-1}$ are in full agreement with previous knowledge about the position and range of the crossover region in this model (which has been widely studied in the literature).   

For the finite-range NSR potential, the curve resulting from the peak position in the fidelity susceptibility 
was compared with the crossover ``phase diagram'' previously obtained in 
Ref.~\onlinecite{and99}, where the borders of the crossover region were defined on the 
basis of the value of the ratio ($k_F \xi_{{\rm pair}}$)  between the pair 
correlation length and the average interparticle distance.
For densities such that the average interparticle distance does not get smaller than the range of the potential, the peak in the fidelity lies as expected in 
the middle of the crossover region, while for higher densities it tends to 
approach the BCS border of the crossover, as defined from the value of 
$k_F \xi_{{\rm pair}}$. 
We have attributed this behaviour to the increasing importance of inter-pair correlations in the extraction of
$\xi_{{\rm pair}}$ from the pair correlation function at high densities.

We have considered finally the attractive Hubbard model. The results in two 
dimensions compared favorably with the corresponding crossover  ``phase diagram'' of Ref.~\onlinecite{and99}. In three dimensions, we have argued that the 
fidelity susceptibility is able to evidence within the superconducting state the traces of an underlying quantum phase transition, which has been found 
previously  for the normal state when the superconducting correlations were artifically suppressed\cite{metzner,capone}.

In summary, from our analysis we conclude that the fidelity susceptibility provides a useful tool to characterize the width and position of the crossover region in crossover problems, which is especially appealing because of  its ``intrinsic'' and quite general character. The definition and location of the crossover region based on the fidelity susceptibility do not depend, in fact, on specific quantities like, for instance, the pair correlation length $\xi_{{\rm pair}}$ in the BCS-BEC crossover, which need to be changed when passing from one crossover problem to another one. Even though our analysis was based on the use of an approximate ground state wave function (the BCS trial wave function), the widely tested reliability of such a  wave function in the context of the BCS-BEC crossover give us confidence in the robustness of our results. It will be however interesting to consider 
in future work alternative methods, like e.g. Quantum Monte Carlo or Density Renormalization Group methods, or alternative crossover problems to confirm our main conclusions and place them in a broader context.           

\acknowledgments
We thank G.C.~Strinati for useful discussions.
Partial support by the Italian MUR under contract PRIN-2007 
``Ultracold Atoms and Novel Quantum Phases'' is acknowledged.

\end{document}